\documentclass[
twocolumn,
]{ceurart}

\usepackage{listings}
\usepackage{graphicx}
\usepackage{booktabs}
\usepackage{lipsum}
\usepackage{xcolor}
\usepackage{tabularx}
\usepackage[capitalise]{cleveref}
\usepackage{appendix}
\usepackage{caption}
\usepackage{subcaption}

\begin{document}

\copyrightyear{2024}
\copyrightclause{Copyright for this paper by its authors.
  Use permitted under Creative Commons License Attribution 4.0
  International (CC BY 4.0).}

\conference{}

\title{Creating Healthy Friction: Determining Stakeholder Requirements of Job Recommendation Explanations}

\author[1]{Roan Schellingerhout}[
    orcid=0000-0002-7388-309X,
    email=roan.schellingerhout@maastrichtuniversity.nl
]
\cormark[1]

\author[1]{Francesco Barile}[
    orcid=0000-0003-4083-8222,
    email=f.barile@maastrichtuniversity.nl
]

\author[1]{Nava Tintarev}[
    orcid=0000-0002-5007-5161,
    email=n.tintarev@maastrichtuniversity.nl
]

\address[1]{%
    Maastricht University, Paul-henri Spaaklaan 1, 6229 EN Maastricht, The Netherlands
}

\cortext[1]{Corresponding author.}

\begin{abstract}
The increased use of information retrieval in recruitment, primarily through \textit{job recommender systems} (\textbf{JRSs}), can have a large impact on job seekers, recruiters, and companies. As a result, such systems have been determined to be high-risk in recent legislature. This requires JRSs to be trustworthy and transparent, allowing stakeholders to understand why specific recommendations were made. To fulfill this requirement, the stakeholders' exact preferences and needs need to be determined. To do so, we evaluated an explainable job recommender system using a realistic, task-based, mixed-design user study ($n=30$) in which stakeholders had to make decisions based on the model's explanations. This mixed-methods evaluation consisted of two objective metrics -- correctness and efficiency, along with three subjective metrics -- trust, transparency, and usefulness. These metrics were evaluated twice per participant, once using real explanations and once using random explanations. The study included a qualitative analysis following a think-aloud protocol while performing tasks adapted to each stakeholder group. We find that providing stakeholders with real explanations does not significantly improve decision-making speed and accuracy. Our results showed a non-significant \textit{trend} for the real explanations to outperform the random ones on perceived trust, usefulness, and transparency of the system for all stakeholder types. We determine that stakeholders benefit more from interacting with explanations as decision support capable of providing healthy friction, rather than as previously-assumed persuasive tools.
\end{abstract}


\maketitle

\section{Introduction}
\label{sec:introduction}
Recommender systems have found their way into many aspects of daily life. Even highly impactful decisions, such as the job that one applies to, and vice versa, the applicants a company considers, are often influenced by so-called job recommender systems. When it comes to such impactful scenarios, blindly relying on algorithms to make the correct decision can be risky and may lead to unintended consequences. As a result, there is a growing demand for explainable artificial intelligence (XAI) within the field of job recommendation, which aims to provide transparent and interpretable insights into the decision-making process of such systems \cite{de2021job,gade2019explainable,zhang2020explainable}.

Most research on XAI, however, focuses on assisting developers and other users with prior knowledge of AI \cite{lopes2022xai, preece2018stakeholders}, with the amount of user-centered research staying rather limited \cite{gutierrez2019explaining, millecamp2019explain, szymanski2022designing}. Considering that the overwhelming majority of users of recommender systems tend to be lay users, such in-depth, technical, and often complicated explanations offer little value. Therefore, it is crucial to design explanations in such a way that they are accessible not just to AI experts, but also to everyday users with different levels of expertise. Within job recommender systems, the everyday users are threefold: \textit{candidates} - those looking for a job; \textit{recruiters} - those whose job it is to match candidates to vacancies; and \textit{company representatives} - those who are responsible for hiring in companies. Considering these stakeholders all perform different tasks, their explanation requirements also tend to differ, making tailor-made solutions and compromises necessary. Previous work investigated the preferences related to three explanation types (textual, bar chart, and graph-based) of each stakeholder type, providing insights useful for the design of such tailor-made solutions \cite{schellingerhout2023codesign}. In this paper, we aim to determine the actual benefit those solutions offer to the different stakeholders. As a result, this paper aims to answer the following research question: 

\textit{RQ: To what degree does the combination of multiple explanation types assist job recommender system stakeholders in their daily tasks?}
To answer this question, we conducted a pre-registered\footnote{\url{https://osf.io/5nuk9}}
mixed design user study ($n=30$), wherein participants from each stakeholder type interacted with a recommendation environment in which they could combine multiple explanation types pertaining to predicted matches between candidates and vacancies, or choose to look at the different types of explanations independently. Using this environment, we explored the following sub-questions:

\begin{itemize}
    \item[SQ1:] To what extent do the explanations assist the stakeholders in their decision-making process?
    \item[SQ2:] What is the impact of different explanation components (textual, bar chart, and graph-based) on the stakeholders' understanding of the explanation?
    \item[SQ3:] How can the explanations be improved to better assist the different stakeholders?
\end{itemize}

Our results show that all three stakeholder types find the inclusion of explanations in job recommender systems beneficial. However, they still largely rely on their knowledge and intuition, often disregarding the explanations, or interpreting them differently than intended by the model. Furthermore, based on our thematic analysis, we find that candidates and recruiters prefer text-based explanations, while company representatives lean more toward visualizations. All stakeholders agreed that the explanations could be improved by showing a more direct relation to the CV/vacancy for which recommendations were made, e.g., by reiterating exact phrases in the explanation, and showing how those were incorporated. How different features contributed to the recommendation (positively/negatively, and to what extent) should also be made explicit, to minimize the risk of misinterpretations. Therefore, we determine that a focus on decision support over persuasion is preferable. 

\section{Related Work and Hypotheses}
\label{sec:rel_work}
As determined by the European Union in the AI act \cite{EU2021ai}, the usage of AI in recruitment can be considered a high-risk scenario. Due to the large impact that career choices can have on individuals' lives, as well as the fact that recruitment often deals with large amounts of sensitive data, job recommender systems require a more tailored approach compared to less impactful recommender systems (e.g., music recommenders) \cite{mashayekhi2022challenge}. However, current state-of-the-art approaches often fail to make use of such tailored approaches, causing aspects such as explainability to be largely ignored in current literature \cite{de2021job}. 
In this section, we provide an overview of the current works of explainability in job recommendation, both for experts and lay users. Additionally, we formulate hypotheses for our research questions based on existing literature. 

\subsection{Explainable job recommendation}
While a number of previous works have incorporated explainability within their JRSs, the explanations often have limited expressive value or were not the main focus of the system \cite{le2019towards,Upadhyay2021,yildirim2021bideepfm}. Even when explainability has been included, authors usually fail to consider all stakeholders, tailoring the explanations to only one group (e.g., developers or users only) \cite{gutierrez2019explaining}. Furthermore, explanations are often solely evaluated anecdotally, leaving their quality up for debate \cite{nauta2022anecdotal}. One could argue that easy-to-understand explainability should be at the core of the models' design in a high-risk, multi-stakeholder domain such as recruitment. Previous research, however, often does not explicitly consider the understandability of their explanations: while their models can technically explain some part of their predictions, the explanations tend to be unintuitive and/or limited, either staying too vague \cite{le2019towards,Upadhyay2021} or being hard to understand \cite{yildirim2021bideepfm} for the intended users. Furthermore, baselines are rarely used for evaluating the effect of explanations, leaving their actual added benefit up for debate. It has been shown that lay users may positively evaluate explanations, even when they do not properly understand them 
\cite{szymanski2021visual}. By comparing two explanations (in our study, a random baseline vs a real explanation) in the same environment, it is possible to determine whether explanations \emph{actually} add value for the users. Although explanations should always be available \cite{EU2021ai, EuropeanParliament2016a}, they will not necessarily always be useful \cite{burke2023closing}. Explanations have been shown to be mostly used whenever the user finds themselves in contention (e.g., whenever they disagree with the recommendations or do not find any of the items suitable) \cite{vasconcelos2023explanations}. As a result, these difficult choices are likely to be alleviated by proper explanations, allowing the users to make the correct decision more quickly and often. Thus, whenever the explanation helps the user make a decision, it will also improve their view of the system as a whole. This leads to the following hypotheses for SQ1 (\textit{To what extent do the explanations assist the stakeholders in their decision-making process?}):

\begin{description}
    \item[H1a:] When provided with the real explanations, participants will be able to find matches more quickly, and make the correct decision more often, compared to when they are provided the random explanations. 
    \item[H1b:] Participants will respond more positively to a recommendation environment that includes the real explanations than to a similar one that includes random explanations. This will improve metrics such as perceived trust, transparency, and usefulness. 
\end{description}

\subsection{Explanations for lay users}
When dealing with users with limited AI knowledge (e.g., recruiters, job seekers, and most company representatives), having clear, straightforward explanations is crucial \cite{schellingerhout2022explainable}. While explainability methods such as feature attribution maps (usually in the form of bar charts) can be sufficient for AI experts to get a better understanding of a model, this is not necessarily the case for lay users. Although such `technical' explanations often \emph{look} intuitive, they can be deceptive by giving the users a false sense of understanding \cite{szymanski2021visual}. In another study, specifically on job recommendations, textual explanations were found to be preferred by the majority of lay users \cite{schellingerhout2023codesign}, but those take additional time to read and understand, limiting their real-world usability. On the other hand, visual explanations tend to include more detail, allowing more experienced users (such as company representatives) to get more value from them. As a result, hybrid combinations of explanation interfaces can be used to make the explanations feel accessible, while still being sufficiently comprehensive. Even when using hybrid combinations, however, unique characteristics of each stakeholder type still play a role. We expect the same preferences to be indicated in our study, as we conduct our experiment in a similar (but more realistic) setting. This leads us to formulate the following hypothesis for SQ2 (\textit{What is the impact of different explanation components (textual, bar chart, and graph-based) on the stakeholders' understanding of the explanation?}):

\begin{description}
    \item[H2:] Candidates and recruiters will mainly use textual explanations to understand the recommendations, while company representatives prefer graph-based explanations. The bar chart will be considered useful as a supportive tool, but will be insufficient as a sole explanation by all stakeholders. 
\end{description}

\begin{table*}[!t]
\begin{tabular}{l|l|l|l}
\toprule
\textbf{}          & \textbf{Age}                     & \textbf{Gender} & \textbf{Industry}                               \\ \hline
\textbf{Candidate} & $M = 26.6, SD=10.08, h=56, l=22$ & 6M, 3F, 1X      & IT, insurance, sociology, finance, law, etc.    \\
\textbf{Company}   & $M = 47.7, SD=11.52, h=69, l=32$ & 5M, 5F          & HR, IT, accounting, education, finance, etc.    \\
\textbf{Recruiter} & $M = 38.0, SD=9.79, h=52, l=24$  & 4M, 6F          & HR, IT, service industry, marketing, etc. \\
\bottomrule
\end{tabular}
\caption{The descriptives of the participants per stakeholder type. Each stakeholder type is represented by 10 participants.}
\vspace{-1em}
\label{tab:descriptives}
\end{table*}

\subsection{Implementing lay user explanations}
When trying to explain recommendations to lay users, multiple design factors need to be considered. Not just the way in which explanations are presented, but also how they are generated, should be carefully taken into account beforehand. Previous works have shown that model-agnostic explainability methods, such as LIME \cite{ribeiro2016should} and SHAP \cite{lundberg2017unified}, can fall short when trying to support non-expert users \cite{chromik2021think,jesus2021can}. Common feature attribution methods, such as LIME, tend to provide limited expressiveness (i.e., they only show the extent to which different features contributed to the prediction, but do not include any sort of interaction or higher-level relations between the features). On the contrary, model-intrinsic methods, such as attention \cite{vaswani2017attention} and integrated gradients \cite{sundararajan2017axiomatic}, can be quite intuitive, even to people with less expertise \cite{schellingerhout2022explainable}. Additionally, such methods lend themselves to the use of graph-based models, which can use knowledge graphs to incorporate additional expressiveness by actually including such high-level relations between features \cite{velivckovic2017graph}. However, importance weights like attention and integrated gradients are limited to the range of [0, 1], even when a feature contributes ``negatively" to a prediction (e.g., a feature that had a very strong, negative impact on the prediction, could still have an attention value of 0.9). This is likely to be considered confusing by some users, as such features should intuitively be given a negative score \cite{teufel2023megan}. 
For recommendations specifically, items are often presented in a list. Therefore, the need can arise to know why a specific item was rated higher than others. For example, when users consider the second-highest rated item to be the best match, they might be interested to see what ultimately caused the model to put another item over their preferred choice \cite{lyu2023listwise}. As a result, we formulate the following hypotheses for SQ3 (\textit{How can the explanations be improved to better assist the different stakeholders?}):

\begin{description}   
    \item[H3a:] The explanations frame the model's reasoning in a `positive' way, which could be perceived as confusing, lowering the usefulness and transparency of - and trust in - the model. All stakeholders will therefore benefit from additionally including negative attention weights. 
    \item[H3b:] There are no comparative (i.e., list-wise) explanations available for the list of recommendations. An additional explanation that explains why recommendation X was ranked higher than Y and Z will therefore be desirable for all stakeholder types. 
\end{description}

\subsection{User studies for evaluation}
While some offline evaluation metrics exist for explanations \cite{nauta2022anecdotal}, it is common practice to evaluate explanations in a realistic setting using individuals from the group that is supposed to use the explanations. The subjectivity of user preference generally requires an approach that allows participants to freely express themselves, as forcing participants to choose from pre-determined answers is likely to lack depth during the formative phase of a system. Therefore, studies evaluating user experience and preference of AI-systems often use a combination of think-aloud protocols and (semi-)structured interviews. For instance, \citet{degen2022explain}, conducted interviews with 11 energy engineers to design an explainable system for such highly expert users. Furthermore, \citet{nelson2020patient} used semi-structured interviews to get insights from 48 patients on the use of AI for skin cancer screening. Similarly, \citet{zhu2023not} combined a think-aloud protocol (i.e., having participants say their thoughts out loud during the experiment) with post-test interviews to evaluate the user experience of an AI-based financial advisory system of 24 users with strongly diverse demographics - both in terms of personal characteristics and level of expertise.

\subsection{Contributions}
This paper aims to evaluate explanations for job recommendations in a realistic setting with a sufficient sample of stakeholders. This is done with the aim of addressing the lack of multi-stakeholder focus and evaluation that is present in previous research. Firstly, we create a novel explainable job recommender system that can generate unique explanations for different stakeholders tailored to their specific needs. Furthermore, we evaluate these explanations through a user study in which the participants have to complete a task that mimics their day-to-day activities. Specifically, we compare a combination of different objective and subjective metrics across two settings, one in which participants see genuine explanations, and one in which they see randomized explanations, in an attempt to determine the explanations' real-world impact.

\section{Methodology}

\label{sec:methodology}
\begin{figure*}[!t]
    \centering
    \includegraphics[width=\textwidth]{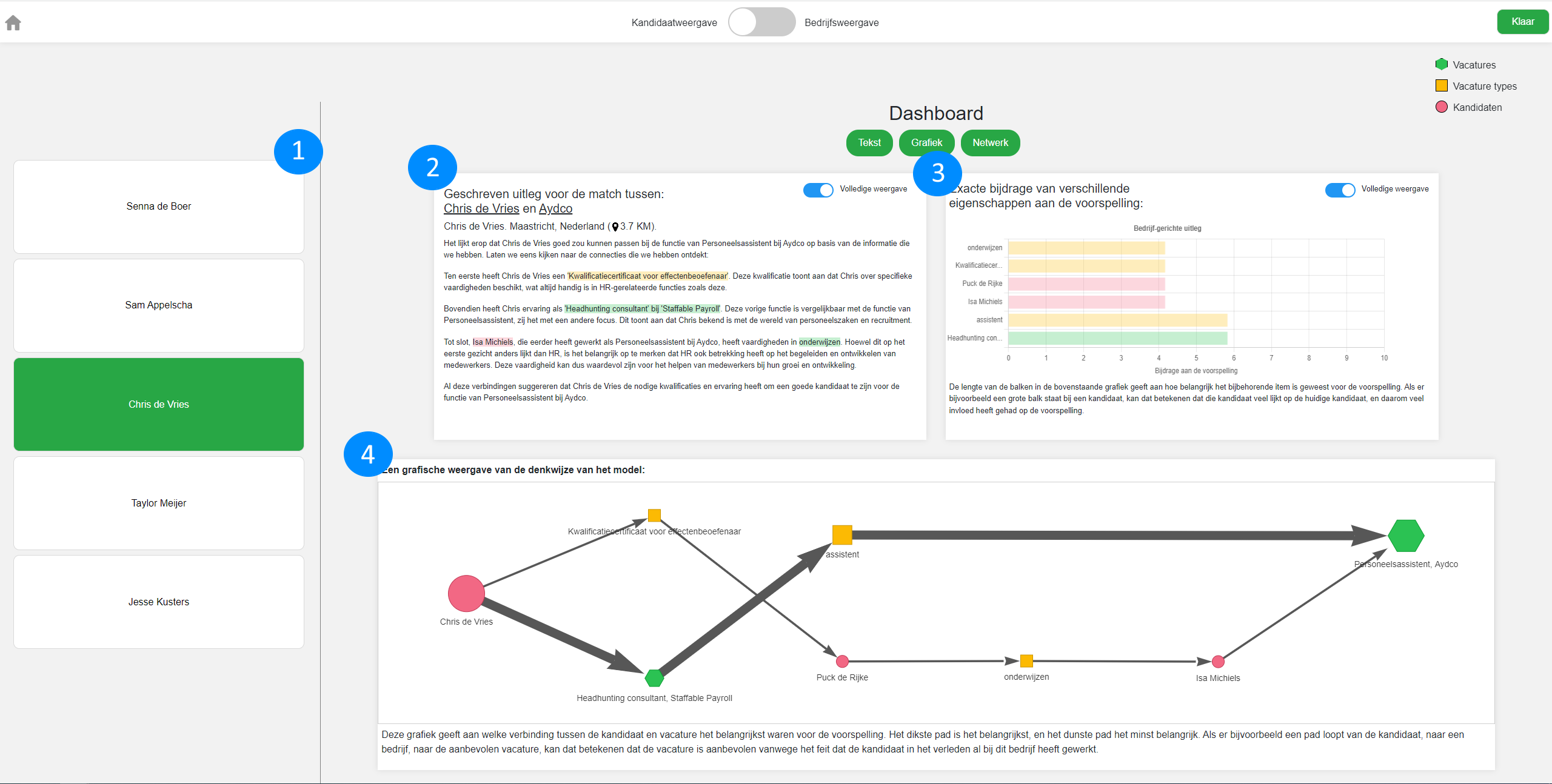}
    \caption{The interface of the online environment with which the participants interacted. In this screenshot, all explanations are enabled. These can individually be toggled based on the user's preference. The web environment uses exclusively Dutch text, as the interviewees were all native Dutch speakers. The environment consists of the following components: (1) the list of recommended items, which were presented in a randomized order (i.e., the top item was not necessarily the best match); (2) the textual explanation; (3) the bar chart explanation; (4) the graph-based explanation. This example shows a \textit{real} explanation. For an example of a \textit{random} explanation, see \cref{app:A}.
    }
    \label{fig:interface}
\end{figure*}

To answer our research question, we performed a pre-registered user study. We created an online environment that allows the different stakeholders to perform tasks that are similar to their day-to-day tasks, e.g., looking for suitable candidates, finding interesting vacancies, or matching candidates to vacancies. An overview of the environment can be seen in \cref{fig:interface}. A total of 10 participants from each stakeholder type were asked to use the environment, for a total sample of 30 individuals. Previous works utilizing similar approaches showed that, due to the quality and amount of data collected through qualitative user studies, such a sample size is sufficient \cite{degen2022explain,morse2000determining,nelson2020patient,zhu2023not}. 
The participants were recruited from a wide range of backgrounds (e.g., area of expertise, age, gender identity, etc.) to mitigate possible biases from arising. Participants were gathered in two ways: 
through personal and professional networks, as well as in collaboration with \textit{Randstad N.V.}\footnote{\url{https://www.randstad.nl/}} (Randstad), the world's largest recruitment agency. 
They were asked to participate en masse over e-mail and were provided an information letter including details on what the research would entail. Our final sample consisted of 14 women, 15 men, and 1 non-binary person with an average age of 37.4 years ($SD = 13.486, l=22, h=69$). The participants had widely varying backgrounds (e.g., IT, HR, finance, sociology, law, marketing), leading to levels of expertise w.r.t. AI ranging from no knowledge whatsoever to a Master's degree in a related field. A full overview of the descriptives per stakeholder type can be seen in \cref{tab:descriptives}. Both qualitative and quantitative data were gathered during the experiment through a previously validated (semi-structured) interview guide \cite{schellingerhout2023codesign}. This allowed participants to freely speak their minds, while also allowing for statistical analyses to be performed.

\subsection{Procedure}
Participants were given the choice to conduct the experiment online or in person. All but one participant preferred to participate online; as a result, 29 out of the 30 interviews were done in a video call wherein the participant shared their screen to allow for active monitoring of their actions. After being sent a link to the online environment, agreeing to a consent form, and filling in their demographics, each participant was asked to perform a task related to their specific stakeholder type: (i) for candidates, this translated to finding the most interesting vacancy from a given list of recommendations; (ii) company representatives were tasked to find the most suitable candidate for an existing vacancy based on a list of recommendations; (iii) recruiters, on the other hand, had to attempt to find the best match between a vacancy and a list of possible candidates. 

These recommendations were based on a third-party dataset (\cref{sec:model}) and were therefore not related to the participants themselves. Considering the dataset we used did not contain written CVs (but only semi-structured lists of work history and skills) or any personal data, we synthetically generated CVs based on skills and work histories using ChatGPT-3.5-turbo. During the experiment, participants had to roleplay as an imaginary candidate/company on whose behalf they were asked to make a decision. The synthetic data was not used during the training process and was only used to allow participants to more easily take on the identity of the entity on whose behalf they were judging, as a written CV is more user-friendly and realistic than an unstructured list of work experience and skills. Still, the information within the CV was the same as that within the list, complemented only by information like company names, years of employment, and more verbose descriptions. To ease the process of empathizing with the CVs and job listings, we gave clear instructions to the users so that they did not base their selections on personal preferences, but on relevance to the CV/job posting instead.\footnote{The exact instructions, as well as implementations and model parameters, are available on \url{https://github.com/Roan-Schellingerhout/evaluating_job_recommendations}}

The aforementioned tasks were all executed within the same environment (\cref{fig:interface}) only differing in what data was being shown to the participants. The participants were allowed to inspect all the recommended items and explanations at their own pace and were not instructed to focus on any given explanation type. They were asked to talk aloud during their task to allow us to get an understanding of how the environment was initially perceived, and what aspects the participants found most interesting. Most participants were able to quickly browse through the five recommended items, causing the selection process to take just over 5 minutes on average. The participants performed their allocated task twice - once with randomized explanations, and once with explanations generated by an explainable graph neural network (eGNN) (\cref{sec:model}) - in random order to minimize learning effects. During both iterations, the participants were monitored to evaluate objective metrics, such as their efficiency and the correctness of their decisions. After each repetition, the participants were interviewed to collect ratings of the environment (by asking participants how they would rate certain aspects on a scale from 1 to 10) and determine where improvements could be made (by probing for more information on why they gave said ratings). In total, the entire process took approximately 30 minutes per participant. Participating candidates were paid the equivalent of Dutch minimum wage (i.e., just over 5 euros) in the form of a bol.com (a popular Dutch online retailer, similar to, e.g., Amazon) gift card for their time. Considering recruiters and company representatives participated during working hours, they were not compensated for their time beyond their regular payroll.

\subsection{Model and data}
\label{sec:model}
The heterogeneous explainable graph neural network (eGNN) used to generate the recommendations was implemented using PyTorch geometric \cite{Fey/Lenssen/2019}. Its architecture builds upon existing model designs \cite{brody2021attentive, shi2020masked, Wang2019, teufel2023megan}, but is altered to allow for heterogeneous data to be considered, while also generating separate predictions and explanations for both the user (candidate) and provider (company). Our implementation consisted of a node and edge embedding layer for both textual (based on DPR \cite{karpukhin2020dense}) and categorical data. Textual data was pre-tokenized to adhere to PyTorch Geometric's data conventions. Then, during inference, the tokens were retrieved, embedded, and readded to the graph. Categorical data was embedded using one-hot encoding. After having embedded the non-numerical data, the entire graph was fed into a general sub-graph embedding layer (using a GATv2conv \cite{brody2021attentive}). The output of this layer was a generic embedding of the entire sub-graph as a whole, which was then fed into two parallel stakeholder-specific scoring layers (using Graph Transformers \cite{shi2020masked}). Both parallel stakeholder-specific layers provided a `matching score' based on the sub-graph embedding as their outputs, which were then combined and fed into a linear layer to make a final prediction. The model was trained on a public job recommendation dataset provided by Zhaopin, China's largest online recruitment platform.\footnote{\url{https://www.zhaopin.com/}} This publicly available dataset contains (i) 4.5 thousand job seekers, who are represented by features such as their age, education, experience, and desires (e.g., preferred city and industry); (ii) 4.78 million job postings, which contain information on the specific job, as well as general details of the company; and (iii) 700 thousand recorded interactions between the two, which consist of four stages: no interaction, browsed (either party showed interest by looking at the other's CV/posting online), delivered (the parties were presented to each other), and satisfied (the parties were actually matched up). To use these labels as ground truth values in the user experiment, we considered the highest-rated item (candidate or job) in the list to be the `correct' answer, with the second highest-rated item to be the second-best answer, etc. making sure there were no ties for the highest-rated position. 

We converted this tabular dataset to a knowledge graph using a manually defined ontology.\footnote{The ontology can be found in our \href{https://github.com/Roan-Schellingerhout/evaluating_job_recommendations}{GitHub Repository}} The ontology was created based on the relations between the available features in the dataset. We converted feature types to edges, and feature values to nodes. E.g., if a candidate had the value `transport' for their `current industry' feature, the resulting triple would be (candidate, worksInIndustry, transport). This allowed for previously non-existing connections between candidates and vacancies to arise (e.g., a path from a candidate to a vacancy based on the candidate sharing a common skill with another person who had previously fulfilled the position). This approach led to a final knowledge graph consisting of over 280 thousand nodes and nearly 1.6 million edges, with every node having 5.6 neighbors on average.

By then creating sub-graphs from this knowledge graph through the use of a k-random walk algorithm ($k=7$, 50 walks per match)\cite{lovasz1993random} between job seekers and vacancies, we created a graph ranking dataset. This was done offline and before the experiment was conducted. We trained the eGNN on this newly created dataset by performing a grid search of hyperparameters (learning rate, embedding sizes, attention heads, etc.) on a randomly-sampled training set, consisting of 80\% of the data. The different configurations were evaluated on a validation set, consisting of 10\% of the data. The last 10\% of the data were used for a test set, which served to estimate the real-world performance of the model. The version of the model used to generate the explanations for the experiment had a normalized discounted cumulative gain (nDCG) of 0.606. Considering our focus lies on explainability over model performance, we considered this score to be acceptable for our case. 

Explanations were generated using the attention weights of the model, which indicated the importance of the different nodes and edges in the graphs. For the simplified view of the explanations (which was enabled by default, and rarely altered), only the top three paths that had received the highest attention scores were considered. To compare during the experiments, we also generated random explanations, which simply randomized the attention scores provided by the model (while still adding up to 1.0). 

The explanations received from the model were turned into JSON-objects to be used in the web environment and were displayed as weighted graphs using vis.js\footnote{\url{https://visjs.org/}} (\cref{fig:interface} component 4). Furthermore, the graphs were turned into bar charts by summing the incoming attention for each node (\cref{fig:interface} component 3), and into textual explanations (\cref{fig:interface} component 2) by feeding the JSONs into chatGPT \cite{chatgpt}.\footnote{The prompt used to generate the explanations is included in our \href{https://github.com/Roan-Schellingerhout/evaluating_job_recommendations}{GitHub repository}.} This strategy to generate textual explanations has been shown to be apt in previous works \cite{schellingerhout2023codesign, susnjak2023beyond}, given that proper instructions are provided. 

\subsection{Variables}
To determine the objective benefit of adding explanations to the recommendations, we compared multiple metrics that represent different aspects of usability and usefulness of the environment. These metrics were compared for the two \textbf{independent variables} present in our design: 

\begin{description}
    \item[Scenario:] the type of explanation presented to the user (within-subject, categorical: random or real);
    \item[Stakeholder type:] the participant's stakeholder type (between-subject, categorical: candidate, recruiter, or company representative).
\end{description}

\noindent For these settings, we then compare the values of the five \textbf{dependent variables}: 

\begin{description}
    \item[Correctness:] whether the decision of the participant is correct, based on the ground truth values in the data (scale from 0-3, where higher values are more correct);
    \item[Efficiency:] the time in seconds it takes the participants to decide;
    \item[Transparency:] how much the explanation helps the participant understand how the model made the recommendation (scale from 1-10);
    \item[Usefulness:] the extent to which the explanation helps the participant actually make a final decision (scale from 1-10);
    \item[Trust:] the level to which the participant trusts the system to have made the correct recommendation (scale from 1-10).
\end{description}

\noindent The subjective metrics (transparency, usefulness, and trust) were determined using a validated interview guide that allows for the collection of both qualitative and quantitative data \cite{schellingerhout2023codesign}. All metrics were compared between the two possible scenarios (real and random explanation) to determine the added benefit of using explanations. 

\subsection{Analysis}
We tested hypotheses H1a and H1b using the Mann-Whitney U test, using the aforementioned dependent and independent variables ($\alpha = 0.05$). We use this test rather than independent sample t-tests due to the relatively small sample size, which was unlikely to meet the normality assumption. Furthermore, we opted for individual tests over an ANOVA, because we are not interested in the difference in performance between the different stakeholders, as our goal is to find the optimal explanation type for each stakeholder type individually. 
Hence, the tests were performed individually for each stakeholder type. 

For hypotheses H2 and H3a-b, we applied a thematic analysis \cite{braun2012thematic} to analyze the responses to the questions in the aforementioned interview guide (which can be found in \cite{schellingerhout2023codesign} and \cref{app:A}). We tested H2 using the answers to Sections 1 and 5, as well as questions 2.1.1 and 3.3 of the interview guide, which focus on participants' interpretation of the explanation and their individual preferences. The answers to the questions in these sections contain information on what components of the explanations the participants used during their decision-making process, i.e., how impactful the individual explanation components were. Similarly, we tested H3a and H3b using questions 2.1.2, 4.2.1, and Section 5 of the interview guide. These questions mainly focus on the perceived shortcomings of the system that the participants would like to see addressed.
\section{Results}
\label{sec:results}

\paragraph{SQ1: To what extent do the explanations assist the stakeholders in their decision-making process?}


\begin{table*}[!ht]
\centering
\begin{tabularx}{.85\textwidth}{@{\extracolsep{\fill}} rccccc}
\toprule

\textbf{Setting} & \textit{\textbf{Correctness $\uparrow$}} & \textit{\textbf{Efficiency $\downarrow$}} & \textit{\textbf{Transparency $\uparrow$}} & \textit{\textbf{Usefulness $\uparrow$}} & \textit{\textbf{Trust $\uparrow$}} \\ \midrule
                 & \multicolumn{5}{c}{\textbf{Candidate}}                                                                                                                                      \\ \cmidrule(l){2-6} 
\textbf{Random}  & \textbf{1.10} (1.04)            & 262.00 (131.82)                  & 7.60 (1.18)                      & 7.30 (0.78)                    & 7.02 (1.22)              \\
\textbf{Real}    & 0.90 (0.32)                     & \textbf{229.60} (96.99)          & \textbf{7.80} (0.82)             & \textbf{7.55} (0.83)           & \textbf{7.39} (1.18)               \\ \midrule
\textbf{}        & \multicolumn{5}{c}{\textbf{Company}}                                                                                                                                        \\ \cmidrule(l){2-6} 
\textbf{Random}  & \textbf{1.10}  (0.94)           & 538.50 (433.07)                  & 7.05 (1.11)                      & 6.80 (1.14)                    & 6.70 (1.35)               \\
\textbf{Real}    & 0.80 (0.79)                     & \textbf{479.50} (229.59)         & \textbf{7.30} (1.11)             & \textbf{7.05} (1.19)           & \textbf{7.00} (1.25)               \\ \midrule
\textbf{}        & \multicolumn{5}{c}{\textbf{Recruiter}}                                                                                                                                      \\ \cmidrule(l){2-6} 
\textbf{Random}  & \textbf{1.80} (0.60)            & 298.80 (147.42)                   & 6.70 (1.35)                      & 5.85 (1.79)                    & 5.35 (2.05)               \\
\textbf{Real}    & 1.40 (0.97)                     & \textbf{282.30} (195.77)          & \textbf{7.20} (1.25)             & \textbf{6.45} (2.36)           & \textbf{5.55} (1.89)              \\ \midrule
\textbf{}        & \multicolumn{5}{c}{\textbf{Overall}}                                                                                                                                        \\ \cmidrule(l){2-6} 
\textbf{Random}  & \textbf{1.33} (0.94)            & 366.43 (300.97)                   & 7.12 (1.27)                      & 6.65 (1.44)                    & 6.36 (1.74)               \\
\textbf{Real}    & 1.03 (0.76)                     & \textbf{330.47} (207.7)           & \textbf{7.43} (1.07)             & \textbf{7.02} (1.61)           & \textbf{6.65} (1.64)              \\ \bottomrule
\end{tabularx}
\vspace{.1em}
\caption{The average ratings and standard deviations for the random and the real explanation scenarios per stakeholder type. $\uparrow$ indicates a \emph{higher} value is better for a metric, while $\downarrow$ indicates a \emph{lower} value is better. Correctness ranges from 0 to 3, while transparency, usefulness, and trust range from 1 to 10. Efficiency is measured in seconds. The best value per stakeholder type is shown in bold.}
\vspace{-0.25in}
\label{tab:ratings}
\end{table*}

\paragraph{Correctness}
When focusing on \textit{correctness}, we find that participants often did not change their decision in between rounds (only 7 participants switched between rounds), meaning the difference in correctness between the random and real explanations was small, and not significant (companies: $U=42, n=20, p=.579$, candidates: $U=51, n=20, p=1.000$, recruiters: $U=40, n=20, p=.481$). However, the trend was for the random explanations to lead to more correct answers than the genuine ones (c.f., \cref{tab:ratings}). 
Participants often used their own knowledge to come to a conclusion, even if that knowledge did not align with the explanation. For example, the candidate most often selected by company representatives and recruiters had work experience similar to the vacancy. While this experience was also the most important feature for the real explanation, it was of below-average importance in the random explanation. Regardless, the participants still described it as the main argument of the model, even when viewing the random explanation. In other words, they considered the weighted arguments provided by the model based more on their own intuition, rather than the importance prescribed by the model. This decision-making process occurred with a broad range of participants, especially those who were more reluctant to trust the system. 

\paragraph{Efficiency} In terms of \textit{efficiency}, both scenarios performed similarly (companies: $U=53, n=20, p=.853$, candidates: $U=40.5, n=20, p=.481$, recruiters: $U=44, n=20, p=.684$). Regardless of the order in which the scenarios were presented, the one shown second usually led to a faster decision. This shows that, since the participants had already decided in the first round, they only had to confirm this decision in the second round (as the same list of options was presented in both rounds). We conducted a post-hoc analysis to better determine the contribution of order effects on efficiency (\cref{sec:posthoc}). Furthermore, the participants largely indicated that they had become more familiar with the environment the second time, making them more adept and efficient when comprehending the explanations (e.g., P16: \textit{``Yes, well, it does take some time; you need to get into it for a bit. But it's slowly starting to make sense now. Indeed, it's not something you immediately grasp and say, oh yes, that's how it works"}). 

\paragraph{Transparency} Regarding the perceived \textit{transparency}, we notice that most participants had some difficulty in finding the difference between the two models; largely because the same list of recommendations was presented for both. Recall that participants viewed the same underlying data and only the \textit{importance of different features} changed. However, upon further inspection, participants did determine the genuine explanations to be slightly better at explaining the match, mostly due to them feeling more `sensible' and `descriptive.'
This difference in transparency was, however, not statistically significant (companies: $U=57, n=20, p=.684$, candidates: $U=52, n=20, p=.912$, recruiters: $U=59.5, n=20, p=.481$). 

Participants who used the \textit{graph-based explanation} more, found the difference in edge and node weights provided by the genuine explanation to be useful when mentally parsing the graph. Since the difference between path weights was larger in the real scenario (i.e., in the random setting, most paths had fairly equal weights - roughly $\frac{1}{N}$ with $N$ being the number of paths. Conversely, in the real setting, paths determined to be important by the model had noticeably more weight than the rest - e.g., $>0.9$), they could more quickly determine what arguments the model had determined to be important, after which they could judge if they agreed with those arguments.


\paragraph{Usefulness}
The real explanations also had a positive, but not significant, impact on participants' perceived \textit{usefulness} (companies: $U=55, n=20, p=.739$, candidates: $U=56.5, n=20, p=.631$, recruiters: $U=63, n=20, p=.353$). In both the random and real scenarios, participants indicated the explanations as being helpful as a push in the right direction, but insufficiently detailed to base an actual decision on. Participants who used the textual explanation most found very little difference in the degree to which the explanations helped them make a decision. This was caused by the fact that both texts included the same arguments (because both were based on the same data), mainly differing in different aspects being described as more or less important. However, this difference was quite nuanced, presenting itself in subtle differences of phrasing (e.g., ``somewhat important" vs. ``very important"), causing it to not be noticed much. In the bar chart and graph, the difference in perceived usefulness was more noticeable, and participants who focused mainly on those explanation types rated the real explanation as slightly more useful. 

\paragraph{Trust}
Regarding the perceived \textit{trust}, the recommendations were rated as slightly more trustworthy when the participants were presented with real explanations because those were perceived as better at explaining why a match was made. However, this increase was not statistically significant for any stakeholder group (companies: $U=55.5, n=20, p=.684$, candidates: $U=59, n=20, p=.529$, recruiters: $U=52.5, n=20, p=.853$). One thing that limited the trust in both scenarios was the inclusion of other candidates in the explanation (e.g., including that a candidate with similar skills to the recommended candidate has also fulfilled a specific vacancy in the past), which often led to confusion and uncertainty.
This was perceived as a weak or nonsensical argument by participants, which decreased the trust they had in the system. Some participants mentioned this could be improved by rephrasing the argument to be more general, rather than referring to individuals, e.g., by mentioning that `similar candidates' fulfilled this vacancy in the past.

\paragraph{SQ2: What is the impact of different explanation components on the stakeholders' understanding of the explanation?}

While trying to find a match, \textbf{recruiters and candidates} strongly preferred the textual explanations, as they found those easiest to work with. Especially those without a `technical' background were reluctant when using the visual explanations, as those required participants to compare and evaluate different numbers. Some participants even ignored the visualizations entirely, as they found the textual explanations to be sufficient. 
\textbf{Company representatives} indicated that the graph-based explanation was helpful, but the opinion was split, and understanding how to read the graph was often reliant on reading the text as well. The complexity of the graphs was exacerbated by overlapping edges, which added an additional layer of difficulty (as they required extra effort to be understood properly). However, once the company representatives had familiarized themselves with the environment, most indicated that they could see themselves using the graph a lot more in the future, as it did give them a quick overview of the connections between the vacancy and candidates. Some participants mentioned the graph-based explanation could be made more user-friendly by relating it more directly to the vacancy at hand, 
e.g., if they could customize the outgoing edges of the vacancy, limiting the graph to connections directly related to what they considered the most important aspects of the vacancy. 
The bar chart was not used much by most stakeholders, only being used actively by a handful of participants, as it provided too little context to substantiate a decision. Since a lot of the explanations were based on connections between different data points, simply having feature attributions did not paint a sufficiently comprehensive picture. Those who did use the bar chart, used it mostly as a supporting tool that could help them determine what parts of the textual explanation were most important. 

\paragraph{SQ3: How can the explanations be improved to better assist the different stakeholders?}

Most participants indicated that the explanations could be helpful when providing an overview of why things were recommended. However, because the explanations only touched on a limited number of features (i.e., at most three paths in the graph)
they found them hard to trust, as there could be additional factors that they would consider important, but had not been considered by the AI. As a result, participants overwhelmingly indicated that the explanations could be used as a `push in the right direction,' but would need additional verification/exploration to be actually used.
Furthermore, the explanations were indicated to be too `generic' by multiple participants. They indicated that explanations would be more useful if they included explicit references to the CV/vacancy (e.g., ``The vacancy requires two years of experience, which the candidate possesses"). This was especially important for hard requirements (such as minimal education or experience), which should be verified before considering the rest of the recommendation.


\section{Discussion}
\label{sec:discussion}
We now discuss our results relating to the three sub-research questions and their accompanying hypotheses. We also address the ethical concerns and limitations of this work and make recommendations for future research directions. 

\subsection{Assisting in decision-making}
Based on our findings we reject both hypotheses H1a (\textit{real explanations will help participants find matches more quickly, and make the correct decision more often, compared to random explanations}) and H1b (\textit{participants will respond more positively to a recommendation environment that includes the real explanations}). For H1a, we found no evidence that the real explanations allowed participants to make the correct decision more often compared to the random explanations, and we merely found a weak trend that the real explanations enabled participants to decide more quickly. 

There was a large difference in efficiency between the first and second rounds, regardless of which of the two scenarios was presented first. Since the lists of items shown in both scenarios were identical (with only their content changing), the participants simply looked for any large differences that could change their minds, rather than going through the full explanations a second time (sometimes even explicitly stating that they would not need to give the explanations another look). We return to the measurement of efficiency in the post-hoc analysis in Section \ref{sec:posthoc}.

Although we did find some trends in line with H1b, these findings were not statistically significant. We determine that this is in large part caused by the fact that most participants did not actively engage with the explanations. While most participants gravitated to the textual explanations, they often used these explanations to look for additional information on the candidate and vacancy, rather than to understand \emph{why a match was made}. They would then use the information they had gathered from the explanations to manually decide for themselves, regardless of what was mentioned in the explanations. This behavior was exacerbated by the fact that the difference between scenarios for textual explanations was relatively small, e.g., alterations in phrasing, such as changing from ``very important" to ``with limited impact." These small differences often went unnoticed, causing participants to rely on their own expertise rather than the model's recommendation. Therefore, when using textual explanations to substantiate a recommendation, the phrasing should be precise, strongly stressing to what degree, and in what way, different factors contributed to the recommendation; simply listing arguments seems insufficient.  

\subsubsection{Post-hoc analyses}
\label{sec:posthoc}
We additionally investigated whether the lack of difference in efficiency was influenced by the fact that participants were exposed to both scenarios. To do this, we only consider the values from the first run of each participant, which was randomly selected to be real or random with a 50\% chance). When doing so, we did not find any statistically significant differences in terms of efficiency for any of the stakeholder types ($U=11.0, N=10, p=.914$ for recruiters, $U=8.5, N=10, p=.667$ for candidates, and $U=15.0, N=10, p=.690$ for company representatives). As in the analysis with the full sample, the trend was toward real explanations allowing the participants to decide more quickly. This lends further support to the conclusion that the difference between the two conditions is small and unlikely to lead to large differences in efficiency.

Surprisingly, there was a trend for participants to make incorrect decisions more often when presented with real explanations. To better understand which kinds of mistakes were made, we analyzed the justifications given by participants who incorrectly switched when presented with the real explanation, or correctly switched when presented with the random explanation.

We find that the random explanation was more likely to enable participants to decide using their prior knowledge:

\begin{quote}

\textbf{P23}: ``Yes, because a little more emphasis is placed on his qualifications in accounting and finance, and his experience"    
\end{quote}

In contrast, the real explanations were more likely to steer them into a specific decision:  

\begin{quote}\textbf{P14:} ``At least, because those lines are thicker, they reflect the pattern much more of, hey, what are the core elements? And then I actually see this line compared to these two lines, then I think you see very clearly that connection is much stronger with [incorrect candidate] than it is with [correct candidate]"    
\end{quote}

Once more, this indicates a lack of engagement with the explanations from the participants. As long as there was no large discrepancy between the participants' initial decision and the explanation provided by the model, they would use the explanation to justify their choice, regardless of whether or not that justification was grounded in the content of the explanation. Considering the random explanations did not have any ``strong'' arguments in the sub-graph (i.e., paths with a significantly higher attention value attributed to them), participants were more often able to disregard the model's arguments and use their prior knowledge instead. Therefore, we conclude that \textbf{participants experienced more (healthy) `friction'} when interacting with the real explanations (as those sometimes disagreed with them), while they could nearly always justify their personal reasoning using the random explanations (as those were not as `decisive' with their weights). 
    
\subsection{Explanation components' impact}
Considering the responses to the interview questions, we can accept H2 (\textit{Candidates and recruiters will mainly use textual explanations ... company representatives will gain more from graph-based explanations}). Although company representatives were more split on the graph-based explanations than expected, they still viewed it more positively than the other stakeholders. While some struggled with it initially, they could see its benefits after comparing it to the text. Especially those with more technical backgrounds (engineering, finance, AI, etc.) were quick to master the graphs. As expected, candidates and recruiters stuck mostly to textual explanations, indicating that those were sufficiently expressive while not being overwhelming or intimidating. For some, the texts were sufficient, and those participants did not view the visual explanations at all (three participants in total). This again shows that most participants were more inclined to use the textual explanations to \textit{create a clear image of the situation} rather than actually considering it as an explanation. I.e., they often picked the facts from the textual explanation (e.g., ``this candidate has X work experience, which is relevant to the job") and then `manually' weighed those individual facts to come to a decision, rather than using the model's assigned relevance. 

\subsection{Improving the explanations}
As for H3a (\textit{... All stakeholders will benefit from additionally including negative attention weights}), our findings are in line with our hypothesis: all stakeholders will benefit from additionally including negative attention weights. 
Although not mentioned explicitly by any participant, it is clear that most struggled to make sense of some of the weights in the explanation (e.g., being confused why something they considered irrelevant had a relatively high weight), showing that it was unclear to them that this was a strong argument \emph{against} the match. This was likely exacerbated by the fact that some participants (incorrectly) assumed the five matches presented in the environment were the \emph{top} five recommendations. They, therefore, assumed that all arguments presented by the explanation were meant to convince them of why the item was a good match, rather than possibly explaining why it was not. 

On the other hand, we found no evidence for H3b (\textit{An explanation that explains why recommendation X was ranked higher than Y and Z will be desirable for all stakeholders}), as an additional explanation that justifies why one recommendation was ranked higher than another was not desirable for all stakeholder types within the current interface. The participants sometimes struggled to choose a single best option (candidate or vacancy), but did so anyway by manually evaluating the explanations for the individual items and analyzing the possibilities. Rather than needing clarification of the difference in ratings between two items, the participants used their own reasoning and prior knowledge (e.g., regardless of what the model said, what they would consider to be relevant work experience) to determine which item was most suitable. Considering participants were sometimes already overwhelmed by the amount of information shown in the interface, additionally including list-wise explanations is likely to work counterproductively. As a result, we determine that the focus of XAI systems in high-risk domains should lie more on decision support, rather than persuading users of the model's prediction's correctness \cite{miller2023explainable}. Concretely, this could take the shape of a system wherein the most promising matches are presented in a list, with their drawbacks (the arguments with the most negative values) and benefits (the arguments with the most positive values) clearly listed. 

The need for interactivity within the system was also implicitly stressed multiple times, e.g., by participants wondering why a specific feature was (not) included in the explanation. By allowing users to pick which features they want to evaluate manually, interactivity could be enabled, while simultaneously lowering the risk of information overload that would occur by showing all arguments at once. Therefore, one approach could be to present the CV/vacancy to the user, with the aspects mentioned in the explanations (e.g., work experience deemed relevant, or a lack of proficiency in a skill) highlighted and selectable -- ideally even with a clear distinction between hard and soft requirements. Users could then choose which aspects they want to consider in their decision based on both what the model recommended, and what they deem important. This would give users access to all of the data, while allowing them to focus their attention on the features most likely to be important, thereby balancing the need for interactivity, the risk of information overload, and the focus on decision-support over persuasion.

\subsection{Ethical concerns}
This paper was granted ethical approval by the ethical committee of our university\footnote{The Ethical Review Committee Inner City Faculties of Maastricht University (\url{https://www.maastrichtuniversity.nl/ethical-review-committee-inner-city-faculties-ercic})}
before the experiments were conducted. However, some ethical concerns are still present.

Considering the large effects job recommender systems can have on stakeholders' lives (or operations), they should be used with great caution and sufficient security checks in place. The main approach to this is to always have a human in the loop who is responsible for making a final decision (e.g., a recruiter who interprets the model's output and decides whether to accept it). 

This also alleviates the second ethical concern present within the field of JRSs: the fear of being replaced that was expressed by recruiters. While the current system did not perform well enough for recruiters to be concerned for now, they did indicate that research on such systems made them feel somewhat uneasy, as they ultimately worried about being replaced. While legislature such as the EU AI act \cite{EU2021ai} prevents recruiters from being replaced entirely, their fear is still justified, as such systems, when implemented without proper ethical guidelines, could allow companies to hire significantly fewer recruiters. To reassure recruiters, JRS research should focus strongly on solely \textit{supporting} recruiters, rather than doing their job for them. This will simultaneously ensure adherence to the current legislature and relieve recruiters. Without sufficient human recruiters, the field of job recommendation will ultimately stagnate, as ground truth values used to train models are still overwhelmingly human-generated. Without such human inputs, the field will ultimately dig its own grave as models will be trained on (possibly biased and incorrect) outputs of previous models. 

    
\subsection{Limitations and future work}

The examples shown in the environment were manually selected to allow for a larger variety of choices and explanations, e.g., similar/dissimilar items (e.g., vacancies in different industries) and straightforward/complicated explanatory paths (explanation graphs ranged from having 6 to 12 total edges). During this selection process, any explanations of noticeably low quality were discarded. 
However, any limitations in model accuracy (model performance was not our main focus) could also influence the perceived quality of explanations, as explanations for very poor recommendations are unlikely to be perceived well. By performing the manual curation process for the explanations, however, we also ensured that the recommendations shown to the participants were at least of moderate quality; we, therefore, expect explanations generated by a better-performing model to be received similarly. Regardless, our future work will use a better-performing model allowing us to explain well a wider range of recommended items, enabling, for example, \emph{online} testing. 

Similarly, we specifically opted for the use of model-intrinsic explanations, as the use of attention mechanisms allowed us to generate separate candidate- and company-side explanations. However, it would have been interesting to compare these explanations to those generated by existing, state-of-the-art, post-hoc methods, namely SHAP \cite{lundberg2017unified} and LIME \cite{ribeiro2016should}. Considering those methods do not allow the generation of such separate explanations, we did not consider them to be within the scope of this research. Regardless, future work could compare the use of stakeholder-specific to `generic' explanations to evaluate to what extent such specific explanations benefit the stakeholders. 

One aspect that has likely dampened the difference between the real and random scenario, is how the explanations were generated. Although the explanation weights were different in the two scenarios, the data on which the explanation was generated was the same, and the randomization only occurred after a list of recommendations was created. This was done to allow for a fair comparison between the two scenarios (as otherwise the list of recommended items would have changed), but ended up making the difference too small to notice for most. This was further exacerbated by the fact that most participants gravitated towards the textual explanations, in which exact values were not shown. Considering most participants did not pay attention to the exact weights of different components, instead focusing on what \textit{they} considered important, having both scenarios include the same list of features, with only their weights altered, ended up strongly diminishing the noticeable differences between the two settings. Therefore, future work should use a more noticeably different baseline to compare explanations to, e.g., a \textit{fully} randomized explanation. 

Additionally, to assist users in actually making use of all important information within explanations, our next work will attempt to mitigate the aforementioned cognitive bias by trying to present explanations that increase healthy friction \cite{buccinca2021to, Hertwig2017nudging, lorenz2020behavioural, rieger2023nudges}, without being overly coercive. As mentioned, we predict interactive interfaces could be a fitting medium for this; however, such interfaces generally have a higher barrier of entry -- it would therefore be interesting to see if and how users improve (in terms of correctness, trust, etc.) over time when exposed to interactive environments. 

Furthermore, while the company representatives and recruiters could easily identify with the vacancy, some candidates struggled to make decisions on behalf of someone else. This was boosted by the fact that the synthetic candidate they had to roleplay was not customizable. This made it difficult for those with niche backgrounds to determine what would be relevant for such a person. 
In our next steps, we plan to address this by allowing candidates to either use their own data to create recommendations or provide different user profiles which would allow candidates to select a CV most closely related to their personal expertise. Using such a setup, it will be possible to conduct an experiment in a more natural setting, wherein participants would need to make actual decisions where disagreement with the system could be studied in depth.

Lastly, we expect to find similar results when performing our experiment with stakeholders from different high-stakes domains (e.g., medicine, finance). Based on our findings, we predict similar domain experts with little AI knowledge to be likely to show similar behavior, i.e., not actively engaging with the explanations, and instead prioritizing their own expertise. Therefore, we expect the demand for supportive explanations over persuasive explanations to be a general trend, not solely relevant to recruitment. However, as stakeholders' needs and preferences can be highly domain-specific, future research should verify these expectations by creating a similar environment catered to the needs of the aforementioned domains' specific stakeholders. 
\section{Conclusion}
In this paper, we find that job recommender system stakeholders struggle to identify differences in the feature importance of recommendation explanations. 
We determined that this is because most participants struggled to differentiate between the two scenarios. 


We also believe that this difficulty in distinguishing real and random explanations is influenced by their pre-existing expectation and knowledge about the domain. When indicating that they struggled to find a difference between the two settings, participants often mentioned that they attributed more value to the high-level arguments (e.g., \textit{``candidate X had experience in field Y"}) that were being provided by the explanations, rather than the details (e.g., \textit{``this experience had a rather limited impact on the prediction"}) of how much those arguments contributed to the final decision. 
Therefore, we determine that job recommendation explanations are mainly useful when providing users with a general overview of the `big picture.' Current implementations, such as feature attribution and textual explanations tend to be insufficient when providing users with detailed information on how much different features weighed in the prediction. However, such details could be especially useful when trying to alleviate confirmation bias.

We found that focusing on the big picture is especially important for textual explanations, as differences in importance can be hard to relativize. Considering textual explanations were the preferred medium for both recruiters and candidates, future systems should make sure those are implemented in strong accordance with stakeholders' demands. For company representatives, on the other hand, visualizations were more useful. While nuanced differences are easier to distinguish in such explanations, the higher-level arguments being included should still be carefully considered, as simply including all arguments (i.e., showing the entire graph at once) will often lead to information overload. Therefore, regardless of the explanation medium being used, interactive interfaces could assist users by allowing them to view all available data, while minimizing the risk of information overload. 

Furthermore, designing the explanations in such a way that they function as decision-support tools, rather than as persuasive tools to convince users of the model's correctness, would be beneficial for all stakeholders. Providing the explanations as a clear description of the strengths and weaknesses of each of the recommendations will make the explanations more clear and allow users to make up their minds independently. We, therefore, recommend practitioners pivot their focus away from designing persuasive explanations for JRSs, instead focusing on decision-support-oriented approaches.


\clearpage

\appendix

\section{Appendix}
\label{app:A}

\begin{figure*}[!b]
    \centering
    \includegraphics[width=\textwidth]{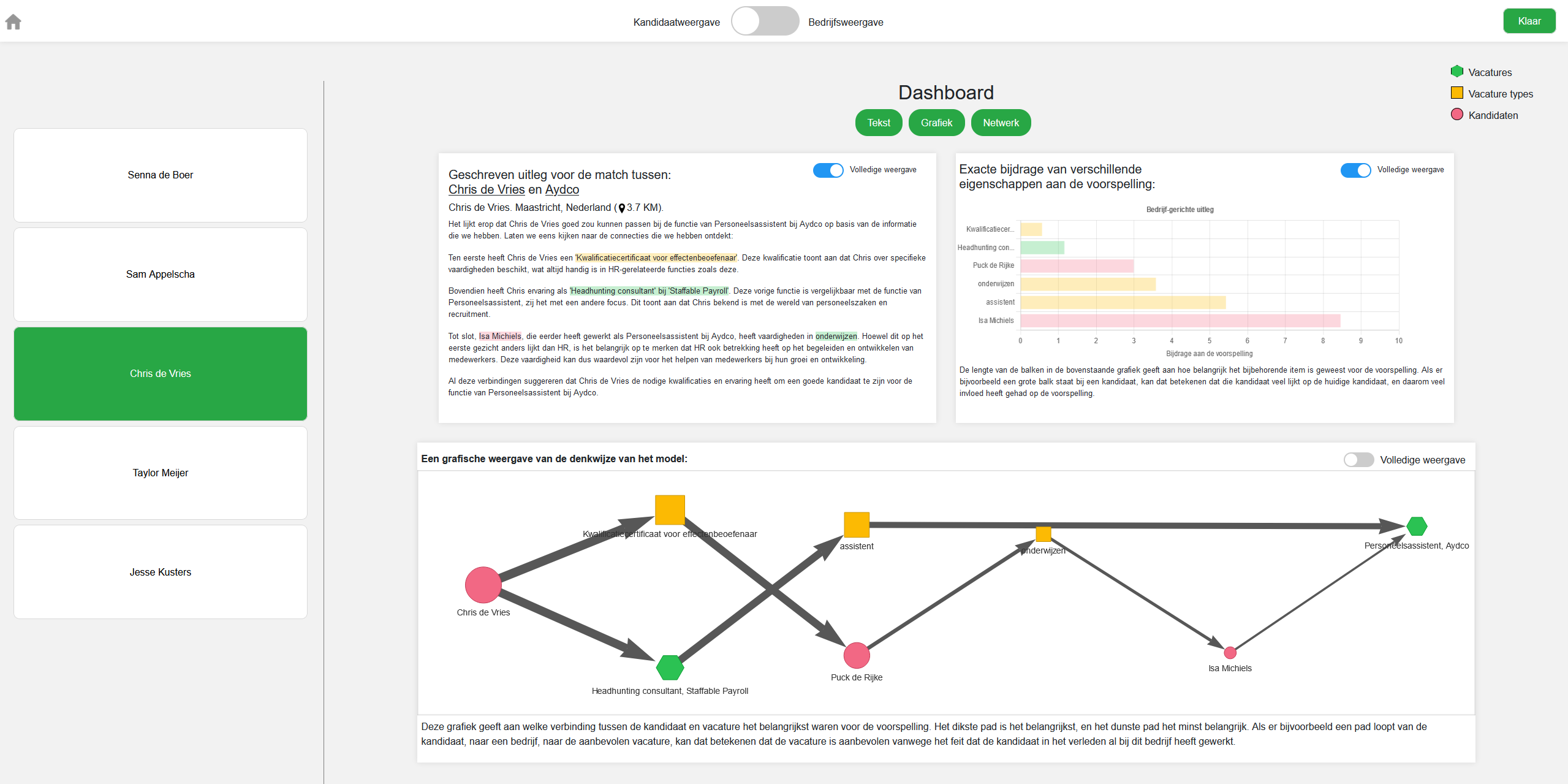}
    \caption{An example of a \textit{random} explanation presented to participants. This explanation relates to the same match as that in \cref{fig:interface}.}
    \label{fig:random_explanation}
\end{figure*}

\cref{fig:random_explanation} shows an example of a random explanation, based on the same candidate as the example in \cref{fig:interface}. The most notable difference occurs in the weights in the graph-based and feature attribution explanations. There, weights differ significantly more than in the real example, with no consistency between connected edges. 

\begin{figure*}[!b]
    \centering
    \includegraphics[width=\textwidth]{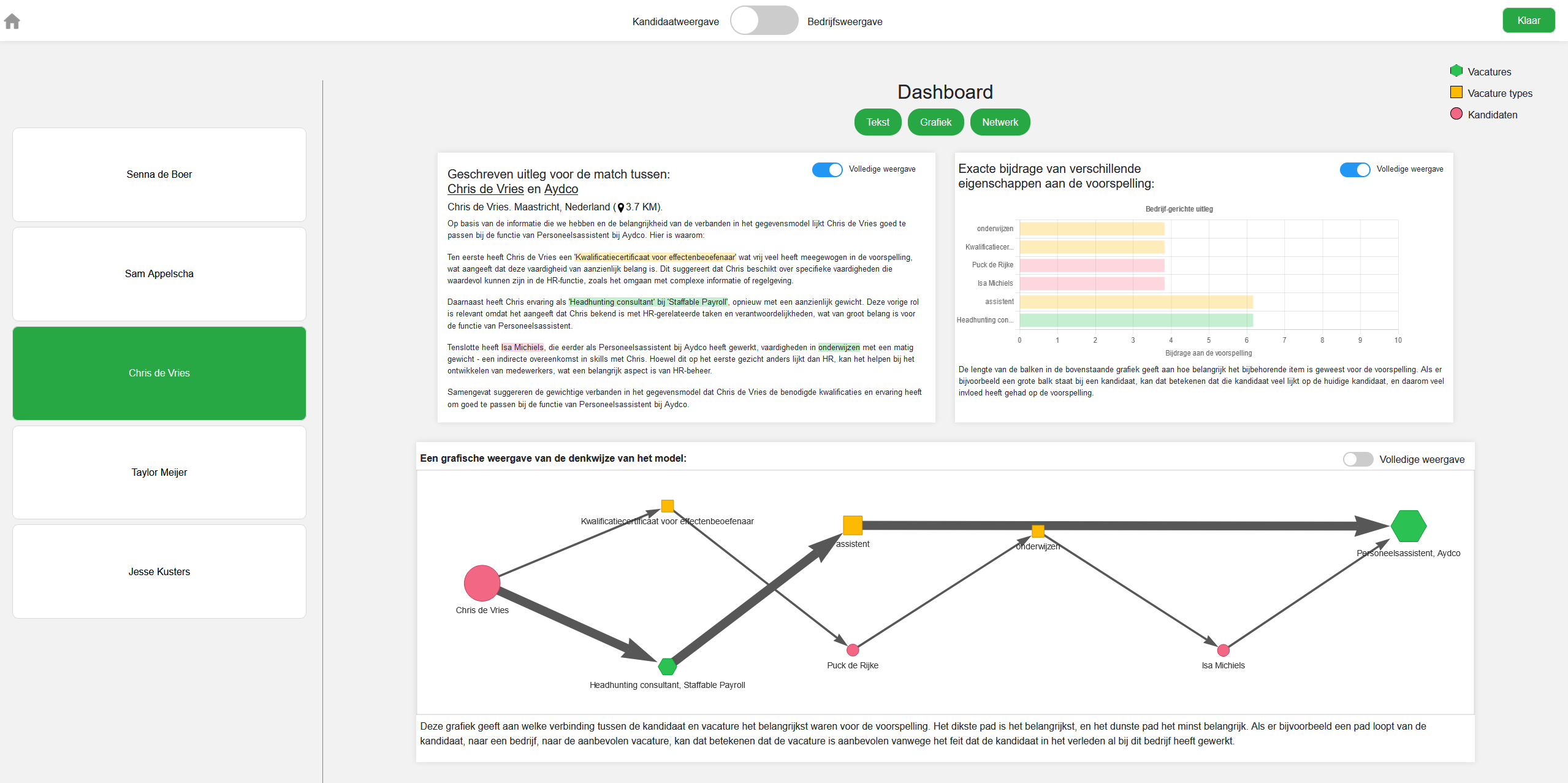}
    \caption{The same explanations as in \cref{fig:interface}, with the indicative numbers removed. This, again, shows the \textit{real} explanations.}
    \label{fig:clean_explanation}
\end{figure*}

\begin{table*}[!t]
\captionsetup{width=\textwidth}
\centering
\begin{tabularx}{\textwidth}{@{}XX>{\raggedright\arraybackslash}p{3.75cm}>{\raggedright\arraybackslash}p{4.5cm}@{}}
\toprule
\textbf{Evaluation Objective} & \textbf{Objective Description} & \textbf{Questions} & \textbf{Probing questions} \\ \midrule

1. Correct interpretation     & To assess whether or not the stakeholder can correctly interpret the explanation. & \begin{enumerate}\item[1.1] What information/features do you think were most important for this prediction? \item[1.2] What was the least important? \item[1.3] How would you put the model's explanation into your own words? \end{enumerate} &         \begin{enumerate}\item[1.1.1] What did you look at to come to that conclusion? \end{enumerate} \\ \midrule

2. Transparency & To determine the explanation's effect on understanding the model's inner workings.                & \begin{enumerate} \item[2.1] Does the explanation help you comprehend why the system gave the recommendation? \end{enumerate} & \begin{enumerate} \item[2.1.1] What components help you specifically? \item[2.1.2] What information is missing that could allow you to get a better understanding of the model's recommendation \end{enumerate} \\ \midrule

3. Usefulness   & To evaluate how useful the explanations are considered to be.                    &     \begin{enumerate} \item[3.1] Does the explanation make sense to you? \item[3.2] Does the explanation help you make a decision? \item[3.3] How could you see yourself using the explanation in your daily work/task? \end{enumerate} & \begin{enumerate} \item[3.1.1] What do you consider sensible (e.g., focus on specific features)? \item[3.1.2] What do you consider insensible? \item[3.2.1] Would you prefer a model with explanations over one without? \end{enumerate} \\ \midrule

4. Trust        & To gauge the explanation's impact on the model's trustworthiness.      &     \begin{enumerate} \item[4.1] Do you think the prediction made by the model is reliable? \item[4.2] If this recommendation was made for you, would you trust the model to have made the right decision? \end{enumerate} & \begin{enumerate} \item[4.2.1] Anything specific that makes you say that (e.g., something makes no sense, or is very similar to how you look at things)? \end{enumerate} \\ \midrule

5. Preference   & To figure out the personal preference of the stakeholder.                        &     \begin{enumerate} \item[5.1] What would you like to see added to the current explanation? \item[5.2] What would you consider to be redundant within this explanation? \end{enumerate} &  \begin{enumerate} \item[5.1.1] Any specific information that is missing? \item[5.1.2] Any functionality that could be useful? \item[5.2.1] Anything that should be removed? \item[5.2.2] Or be made less prevalent?\end{enumerate}             \\ \bottomrule
\end{tabularx}
\caption{The interview guide used to conduct the experiment, based on \cite{schellingerhout2023codesign}.}
\label{tab:interview_guide_updated}
\end{table*}

\cref{tab:interview_guide_updated} contains the interview guide used during the experiment. In addition to these open-ended questions, candidates were also asked to rate the three subjective evaluation metrics (transparency, usefulness, and trust) on a scale from 1-10. 

\end{document}